\documentclass{emulateapj}
\usepackage{rotating}

\begin{document}
\title{Cataclysmic Variables from SDSS III. The Third Year$^{1}$}

\author{Paula Szkody\altaffilmark{2},
Arne Henden\altaffilmark{3,4},
Oliver Fraser\altaffilmark{2},
Nicole Silvestri\altaffilmark{2}, 
John Bochanski\altaffilmark{2},
Michael A. Wolfe\altaffilmark{2},
Marcel Ag\"ueros\altaffilmark{2},
Brian Warner\altaffilmark{5},
Patrick Woudt\altaffilmark{5},
Jonica Tramposch\altaffilmark{2},
Lee Homer\altaffilmark{2},
Gary Schmidt\altaffilmark{6},
Gillian R. Knapp\altaffilmark{7},
Scott F. Anderson\altaffilmark{2},
Kevin Covey\altaffilmark{2},
Hugh Harris\altaffilmark{3},
Suzanne Hawley\altaffilmark{2},
Donald P. Schneider\altaffilmark{8},
Wolfgang Voges\altaffilmark{9},
J. Brinkmann\altaffilmark{10}}
%\affil{email addresses:
%szkody@astro.washington.edu,aah@nofs.navy.mil,fraser@astro.washington.edu,
%nms@astro.washington.edu,bochansk@astro.washington.edu} 
\altaffiltext{1}{Based on 
observations obtained with the Sloan Digital Sky Survey and with the
 Apache Point
Observatory (APO) 3.5m telescope, which are owned and operated by the
Astrophysical Research Consortium (ARC)}
\altaffiltext{2}{Department of Astronomy, University of Washington, Box 351580,
Seattle, WA 98195}
\altaffiltext{3}{US Naval Observatory, Flagstaff Station, P. O. Box 1149,
Flagstaff, AZ 86002-1149}
\altaffiltext{4}{Universities Space Research Association}
\altaffiltext{5}{Department of Astronomy, University of Cape Town, Rondebosch
7700, S. Africa}
\altaffiltext{6}{The University of Arizona, Steward Observatory, Tucson, AZ 85721}
\altaffiltext{7}{Princeton University Observatory, Princeton, NJ 08544}
\altaffiltext{8}{Department of Astronomy and Astrophysics, 525 Davey Laboratory,Penn State University, University Park, PA 16802}
\altaffiltext{9}{Max-Planck-Institut f\"ur Extraterrestrische Physik, Geissenbachstr. 1, D-85741 Garching, Germany}
\altaffiltext{10}{Apache Point Observatory, Box 59, Sunspot, NM 88349}
\begin{abstract}

This paper continues the series that identifies new cataclysmic variables
found in the Sloan Digital Sky Survey. We present 36 cataclysmic variables 
and one possible symbiotic star 
 from Sloan spectra
obtained during 2002, of which 34 are new discoveries, 2 are known
dwarf novae (BC UMa, KS UMa) and one is a known CV identified from the 2dF
survey. The positions, colors and spectra of all 37 systems are presented,
along with follow-up spectroscopic/photometric observations of 10 systems. 
As in the past 2 years of data, the new SDSS systems show a large variety
of characteristics based on their inclination and magnetic fields,
including 3 eclipsing systems, 4 with prominent \ion{He}{2} emission, and 15
systems showing features of the underlying stars.
\end{abstract}

\keywords{binaries: eclipsing --- binaries: spectroscopic --- 
cataclysmic variables --- stars: dwarf novae}

\section{Introduction}

At the present time, the Sloan Digital Sky Survey (SDSS) data have been
released to the public as the Early Data Release (EDR; Stoughton et al. 2002),
Data Release 1 (DR1; Abazajian et al. 2003, 
see http://www.sdss.org), and Data Release 2 (DR2; Abazajian et al. 2004). 
Mining
the database for cataclysmic variables (CVs) has 
resulted in the discovery of 19 new CVs
from the SDSS spectra obtained through 2000 December 31 (Szkody
et al. 2002; Paper I) and an additional 35 from spectra through 
2001 December
31 (Szkody et al. 2003b; Paper II). This paper continues the series that is 
reporting CVs on a
yearly basis with 36 systems (and one possible symbiotic star) that were found among
 the SDSS spectra obtained during the year
2002. Three of these objects were previously known from past X-ray or optical
surveys, whereas the rest are new
discoveries. Follow-up photometric and spectroscopic 
observations of the new systems, providing
 light and radial velocity curves,
 can allow a determination of
their orbital period, which is a fundamental parameter in identifying CVs
(see Warner 1995 for a comprehensive review of all types of CVs).
   
\section{Observations and Reductions}

The details of the SDSS imaging and spectroscopic instrumentation and reductions are
explained in Paper I and in the papers by 
Fukugita et al. (1996), Gunn et al. (1998),
Lupton, Gunn, \& Szalay (1999), York et al. (2000),
 Hogg et al. (2001), Lupton et al. (2001), 
Smith et al. (2002), and Pier et al. (2003).
Briefly, SDSS photometry in 5 filters ($\it{u,g,r,i,z}$) is used to
select objects by color for later spectroscopy in the range of 
3900-6200\AA\ (blue beam) and 5800-9200\AA\ (red
beam) at a resolving power of $\sim$1800. 
The spectra are calibrated for wavelength and flux and 
then classified as stars, galaxies and quasars. As explained in Papers I and II,
cataclysmic variables have spectra obtained through several color selection
algorithms, since their colors overlap with those of hot stars, quasars, white
dwarf and M stars, depending on how much the accretion disk or accretion
column contributes to the optical light over that of the underlying white
dwarf and late-type secondary star.
 The only difference in the procedures used in this paper over
those in the past 2 papers is that we have automated the identification process with
a computer script that finds all spectra with Balmer absorption or
emission lines on a given plate 
(instead of inspecting all spectra on a plate by hand). The resulting spectra
 are then hand-identified as white dwarfs or CVs. As a cross-check on the
accuracy of the computer identification, all the DR1 plates were run through
the algorithm and all the previous CVs identified in Papers 1 and 2 were
recovered.

The list of 36 CVs (and the possible symbiotic) found in SDSS spectra taken in 2002 is presented in Table 1. As in the previous listings,
the magnitudes and colors are from the point-spread function photometry with
no correction for interstellar reddening. To make it easier to locate 
objects in the
spectral database, we give the plate, fiber and MJD of each spectrum in the
Table. For convenience, we will refer to these objects throughout the rest
of this paper as SDSS~Jhhmm, except for 2 objects which have identical
RA coordinates, so we add the first 2 digits of declination to discriminate.  
\begin{figure*}[!htb]
\resizebox{1.0\textwidth}{!}{\rotatebox{0}{\plotone{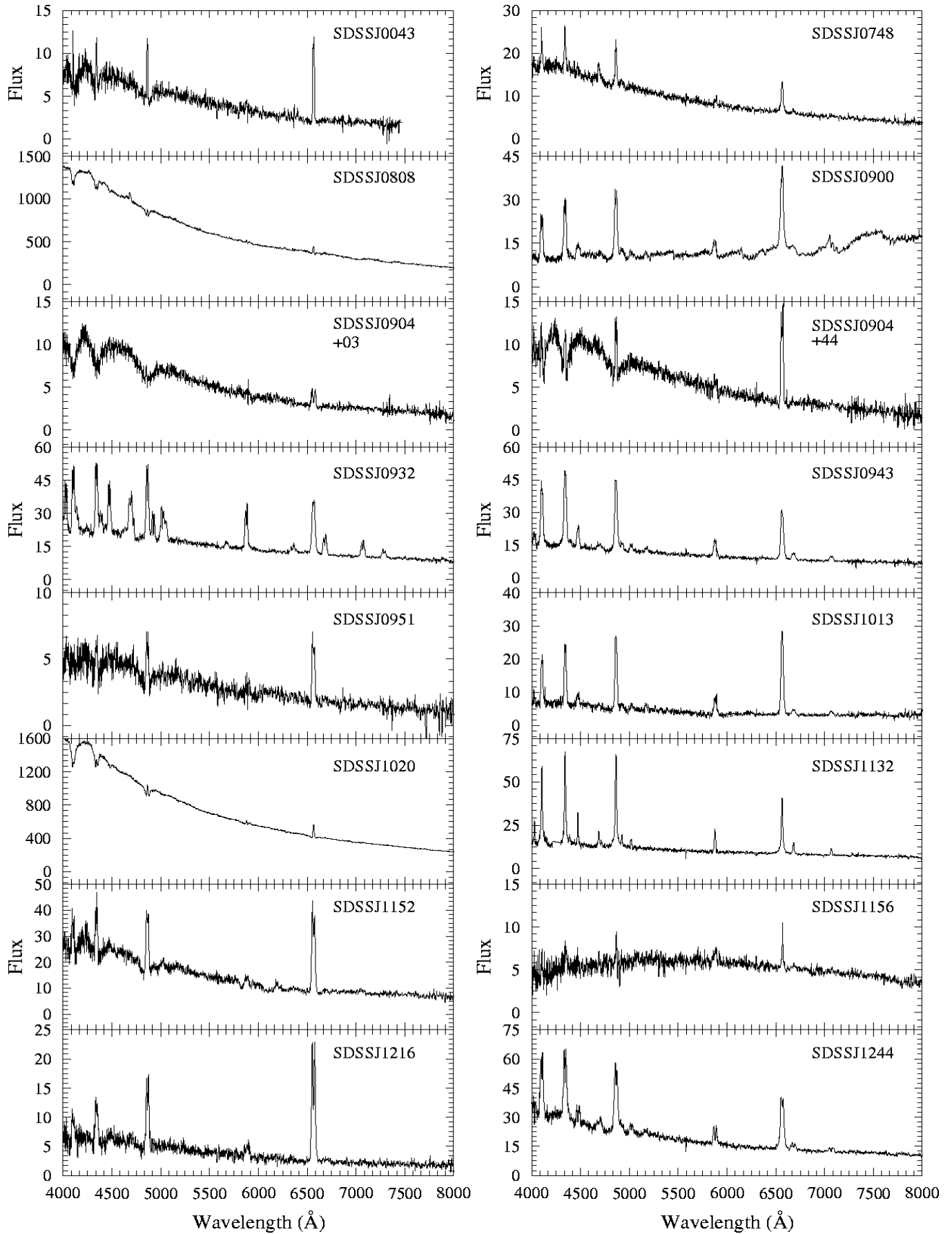}}}
\caption{SDSS spectra of the 36 CVs and one possible 
symbiotic. The flux scale is in 
units of
flux density 10$^{-17}$ ergs cm$^{-2}$ s$^{-1}$ \AA$^{-1}$. The spectral
resolution is about 3\AA.}
\end{figure*}
\begin{figure*}[!htb]
\figurenum {1}
\resizebox{.995\textwidth}{!}{\rotatebox{0}{\plotone{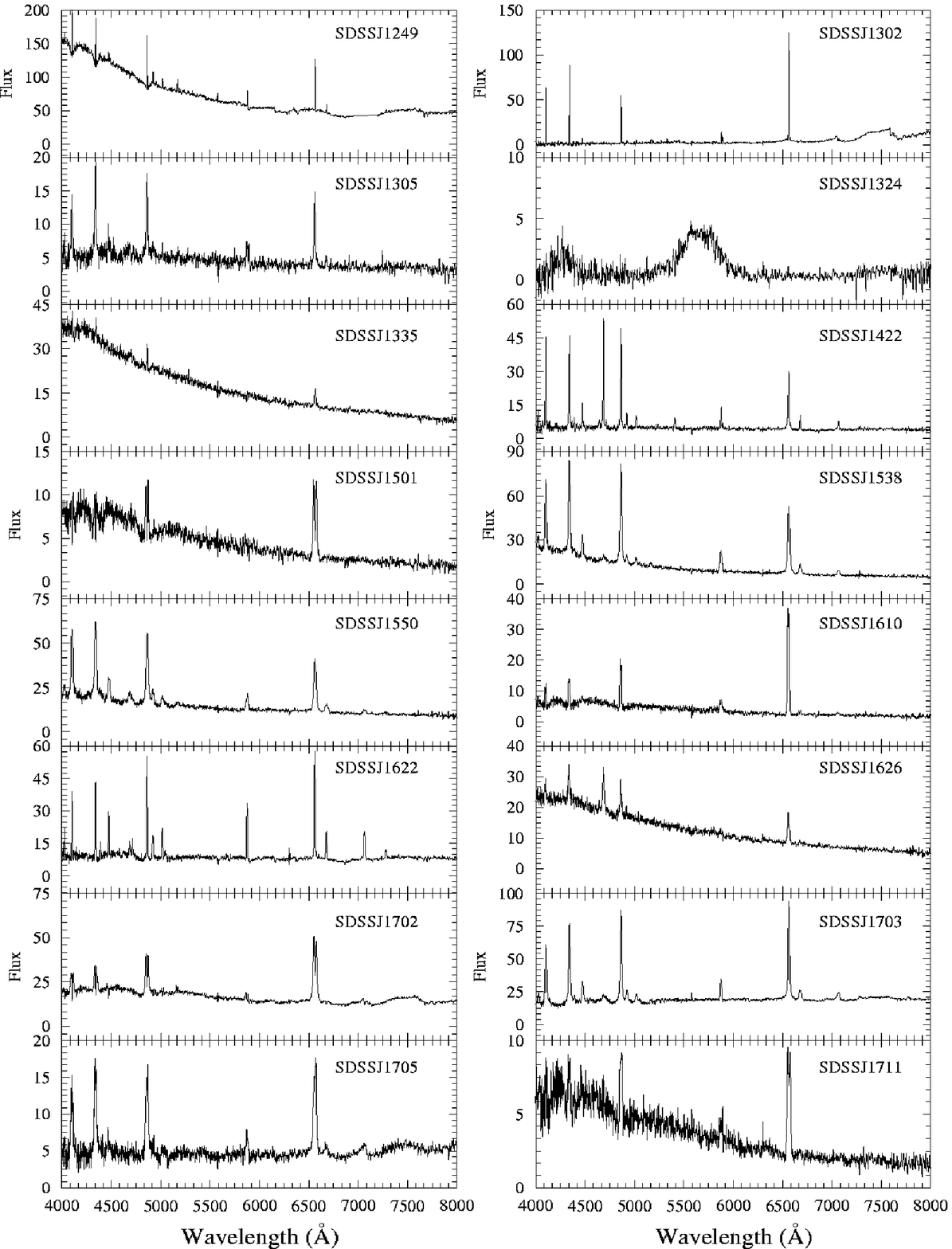}}}
\caption{{\em Continued}}
\end{figure*}
\vspace*{20mm}
\begin{figure*}[!htb]
\figurenum {1}
\resizebox{.99\textwidth}{!}{\rotatebox{0}{\plotone{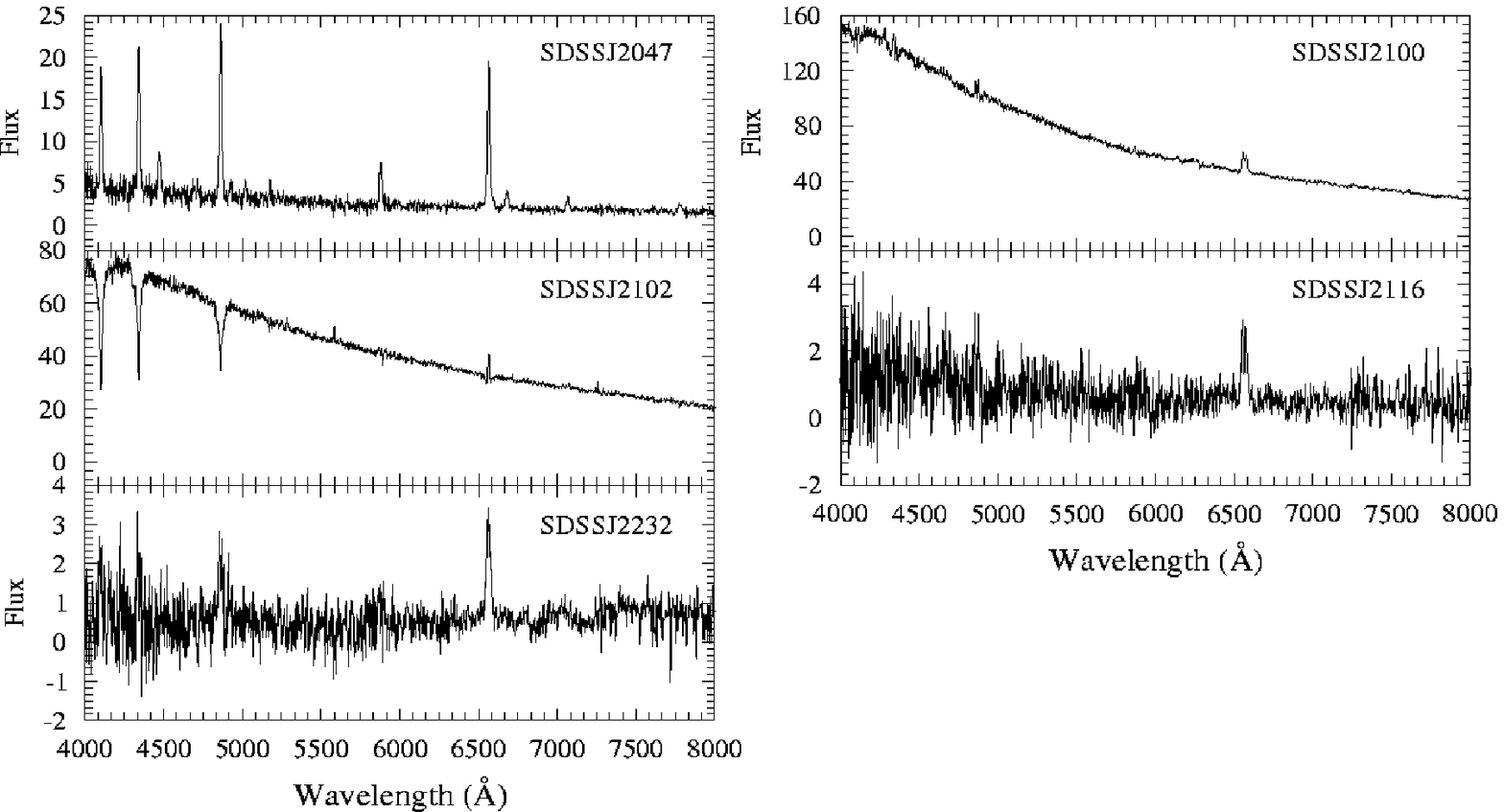}}}
\caption{{\em Continued}}
\end{figure*}

\hspace*{10cm}. We were able to conduct follow-up observations on several systems 
to refine their characteristics (predominantly an estimate of their 
orbital period and whether they 
were at a high enough inclination to be eclipsing). 
Photometry was 
accomplished with the US Naval Observatory Flagstaff Station
(NOFS) 1m telescope
with a 1024 $\times$ 1024 SITe/Tektronix CCD.
Differential
photometry with respect to stars in the field was used to obtain light curves.
With one exception, no filter was used to maximize the
S/N, but 
nights of calibrated all-sky photometry with Landolt standards were used to
calibrate the comparison stars so that the photometry could be placed
onto the Johnson $\it{V}$ magnitude system. For the brightest system (SDSS~J0808),
a $\it{V}$ filter was used for the observation. 

\begin{deluxetable*}{llcrrrrll}
\tablewidth{0pt}
\tablecaption{Summary of CVs with SDSS Spectra in 2002\tablenotemark{a}}
\tablehead{
\colhead{SDSS~J} &  \colhead{MJD-P-F\tablenotemark{b}} & 
\colhead{$g$} & \colhead{$u-g$} & \colhead{$g-r$} & 
\colhead{$r-i$} &
\colhead{$i-z$} & \colhead{$P$ (hr)\tablenotemark{c}} & \colhead{Comments\tablenotemark{d}} }
\startdata

004335.14$-$003729.8 & 52531-1085-175 & 19.84  & 0.37  & 0.01  & --0.23  & 0.18  & 1.3: &\nodata\nl
074813.54$+$290509.2\tablenotemark{e} & 52618-1059-407 & 18.62 & --0.08 & --0.16 & 0.01 & --0.05 & 2.5: &  HeII NP \nl
080846.19$+$313106.0* & 52318-861-352 & 19.43 & --0.29 & 0.68 & 0.62 & 0.40 & \nodata & DN \nl
090016.56$+$430118.2* & 52294-831-435 & 18.88 & --0.22 & 0.68 & 0.69 & 0.49 & 5.3: &\nodata\nl
090403.48$+$035501.2* & 52238-566-380 & 19.24 & 0.26 & --0.09 & --0.18 & 0.04 & 1.4 & ec \nl
090452.09$+$440255.4* & 52294-831-410 & 19.38 & 0.29 & --0.11 & --0.23 & --0.07 & \nodata &\nodata\nl
093249.57$+$472523.0* & 52316-834-431 & 17.81 & 0.22 & --0.03 & --0.08 & 0.05 & 1.7: & HeII \nl
094325.90$+$520128.8* & 52281-768-13 & 16.45 & 0.15 & --0.10 & --0.15 & --0.16 & \nodata &\nodata  \nl
095135.21$+$602939.6* & 52282-770-358 & 20.02 & --0.12 & 0.03 & --0.21 & 0.16 & \nodata &\nodata\nl
101323.64$+$455858.9 & 52614-944-397 & 18.86 & --0.33 & --0.03 & 0.15 & 0.29 & \nodata &\nodata\nl
102026.53$+$530433.1* & 52381-904-147 & 17.44 & --0.08 & 0.24 & --0.04 & 0.28 & 1.6 & KS UMa \nl
113215.50$+$624900.4* & 52319-776-381 & 18.49 & --0.30 & 0.16 & 0.08 & --0.06 & \nodata &\nodata\nl
115215.83$+$491441.9 & 52412-968-280 & 18.51 & 0.14 & --0.10 & --0.09 & 0.31 & 1.5 & BC UMa \nl
115639.48$+$630907.7* & 52320-777-574 & 20.73 & 0.38 & 1.79 & 0.22 & --0.04 & \nodata &\nodata\nl
121607.03$+$052013.9* & 52378-844-423 & 20.12 & 0.21 & 0.12 & --0.29 & 0.00 & \nodata &\nodata\nl
124426.26$+$613514.6* & 52373-781-93 & 18.76 & --0.04 & --0.11 & 0.07 & 0.35 &\nodata  &\nodata\nl
124959.76$+$035726.6* & 52426-847-46 & 16.63 & --0.35 & --0.05 & 0.38 & 0.28 & \nodata &\nodata\nl
130236.21$+$060148.0* & 52439-849-419 & 21.59 & 2.33 & 1.62 & 2.26 & 1.25 &\nodata  & symb-like \nl
130514.73$+$582856.3* & 52410-958-297 & 19.27 & --0.17 & --0.04 & 0.19 & 0.05 &\nodata  &\nodata\nl
132411.57$+$032050.5* & 52342-527-323 & 22.12 & 1.21 & 1.63 & 0.26 & 0.79 & 2.6 & Polar \nl
133540.76$+$050722.7* & 52374-853-498 & 18.46 & --0.02 & --0.17 & --0.16 & --0.20 & \nodata &\nodata\nl
142256.31$-$022108.1* & 52404-918-301 & 19.84 & --0.35 & 0.50 & 0.36 & 0.27 &\nodata  & 2dF, HeII \nl
150137.22$+$550123.4* & 52353-792-44 & 19.40 & 0.24 & 0.02 & --0.21 & 0.01 & 1.3 & ec \nl
153817.35$+$512338.0* & 52401-796-388 & 18.61 & --0.23 & --0.28 & --0.14 & 0.21 & 1.6: &\nodata\nl
155037.27$+$405440.0* & 52468-1053-14 & 18.58 & --0.34 & 0.21 & 0.11 & --0.17 &\nodata  &\nodata\nl
161030.35$+$445901.7* & 52443-814-280 & 19.81 & 0.08 & --0.06 & --0.47 & 0.14 &\nodata  &\nodata\nl
162212.45$+$341147.3 & 52522-1057-203 & 19.15 & 0.19 & 0.43 & 0.30 & 0.40 &\nodata  & HeII \nl
162608.16$+$332827.8 & 52520-1058-433 & 18.38 & 0.08 & --0.21 & --0.21 & --0.07 & \nodata &  HeII \nl
170213.26$+$322954.1* & 52426-973-144 & 17.92 & --0.60 & 0.10 & 0.40 & 0.60 & 2.5 & ec \nl
170324.09$+$320953.2* & 52426-973-97 & 18.17 & --0.77 & 0.56 & 0.36 & 0.56 &\nodata  &\nodata\nl
170542.54$+$313240.8* & 52411-975-400 & 19.67 & 0.12 & 0.48 & 0.48 & 0.25 &\nodata  &\nodata\nl
171145.08$+$301320.0* & 52410-977-402 & 20.25 & 0.20 & 0.05 & --0.09 & --0.09 &\nodata &\nodata \nl
204720.76$+$000007.7* & 52466-982-477 & 19.40 & --0.12 &  0.15 & 0.02 & 0.02 & \nodata &\nodata\nl
210014.12$+$004446.0* & 52442-984-533 & 18.78 & --0.19 & 0.10 & 0.07 & 0.03 & 1.9 & DN \nl
210241.09$-$004408.3* & 52438-984-52 & 17.22 & 1.29 & 0.10 & 0.07 & 0.02 &\nodata  &\nodata\nl
211605.43$+$113407.5* & 52468-729-363 & 15.31 & 0.04 & --0.26 & --0.16 & --0.20 &\nodata &\nodata  \nl
223252.35$+$140353.0 & 52521-738-509 & 17.69 & 0.14 & --0.15 & --0.08 & --0.09 &\nodata  & DN \nl
\enddata
\tablenotetext{a}{Objects marked with asterisk are publicly available in the SDSS DR2}
\tablenotetext{b}{Last 5 digits of Modified Julian Date of Spectrum-Plate-Fiber}
\tablenotetext{c}{Periods marked with : are estimates based on single-night
observations}
\tablenotetext{d}{DN is a dwarf nova, ec is eclipsing, 
NP is not polarized, symb is symbiotic}
\tablenotetext{e}{Object is southeastern star of a close pair}
\end{deluxetable*}

Time-resolved spectroscopy was accomplished at the Apache Point Observatory 
(APO) 3.5m telescope with 
the Double Imaging Spectrograph in high resolution mode (resolution about
3\AA) and a 1.5 arcsec slit. The reductions were identical to those of Papers
I and II. 
Velocities were measured with either the ``e'' (which determines the centroid
of the line) or ``k'' (which fits a Gaussian to the line)
 in the IRAF \footnote{{IRAF (Image
 Reduction and Analysis
Facility) is distributed by the National Optical Astronomy Observatories, which
are operated by AURA,
Inc., under cooperative agreement with the National Science Foundation.}} 
$\it{splot}$ package (for simple or narrow line structure) or with a 
double-Gaussian method (Shafter 1983) for lines with central absorption or
narrow emission components within the broad emission. Least-squares fitting of
sinusoids to the 
velocities was then performed to determine $\gamma$ (systemic velocity), 
K (semi-amplitude), P (orbital period) and $\phi_{0}$ (phase of crossing from
redshift to blueshift). 
Note that since the spectroscopic solutions from APO data are based on a
single night of data, the periods are only estimates which need to be
fully determined with longer datasets. The estimates are designated in Table
1 with a colon following the period.

Circular polarization measurements were obtained for SDSS~J0748, using 
the CCD Spectropolarimeter SPOL with a low resolution grating on the 6.5m MMT at Mt. Hopkins (MH). 
 The dates and types of follow-up observations are summarized in Table 2.

\begin{deluxetable*}{llcccl}
\tablewidth{0pt}
\tablecaption{Follow-up Data}
\tablehead{
\colhead{SDSS~J} & \colhead{UT Date} & \colhead{Site} &
\colhead{Time (UT)} & \colhead{Exp (s)} & \colhead{Data Obtained} }
\startdata
0043 & 2003 Sep 27 & APO & 06:22--09:05 & 900 & 10 spectra \nl
0748 & 2003 Sep 27 & APO & 09:28--12:04 & 600/900 & 10 spectra \nl
0748 & 2003 Nov 1\phn & MMT & 09:52--10:32 & 2400 & spectropolarimetry \nl
0748 & 2004 Jan 28 & NOFS & 02:04--06:16 & 180 & photometry \nl 
0808 & 2003 Nov 27 & APO & 09:04--09:24 & 1200 & 1 spectrum \nl
0808 & 2004 Jan 26 & NOFS & 03:04--12:10 & 90 & V photometry \nl
0900 & 2002 Feb 22 & APO & 02:41--06:41 & 600 & 21 spectra \nl
0900 & 2004 Jan 29 & NOFS & 02:14--11:44 & 180 & photometry \nl
0932 & 2002 Jan 4\phn & APO & 05:38--09:26 & 900 & 14 spectra \nl
1501 & 2003 Jun 25 & NOFS & 04:05--09:22 & 240 &  photometry \nl
1538 & 2002 Jun 26 & APO & 04:54--06:40 & 900/600 & 9 spectra \nl
1702 & 2002 Oct 3\phn & APO & 02:29--03:29 & 600 & 6 spectra \nl
1702 & 2003 Apr 27 & APO & 10:37--11:34 & 600/900 & 4 spectra \nl
1702 & 2003 Jun 26 & APO & 06:52--10:10 & 600 & 16 spectra \nl
1702 & 2003 Jul 3\phn & NOFS & 03:34--10:48 & 180 & photometry \nl
2102 & 2003 Dec 21 & APO & 01:19--02:14 & 600 & 5 spectra \nl
2232 & 2003 Oct 20 & NOFS & 01:33--08:34 & 240 & photometry \nl
\enddata
\end{deluxetable*}

\section{Results}

Figure 1 shows the SDSS spectra for all 37
systems, while 
Table 3 lists the equivalent widths and fluxes of the prominent hydrogen Balmer
and helium emission lines. Using the appearance of the spectra and the follow-up data
when available, we can separate the objects into various categories as
described below. 
\begin{deluxetable}{lrrrrrrrr}
\tablewidth{0pt}
\tablecaption{SDSS Emission Line Fluxes and Equivalent Widths\tablenotemark{a}}
\tablehead{
\colhead{SDSS~J} &  
\multicolumn{2}{c}{H$\beta$} &
\multicolumn{2}{c}{H$\alpha$} &
\multicolumn{2}{c}{He4471} & \multicolumn{2}{c}{HeII4686}\\
\colhead{} &  \colhead{$F$} &
\colhead{EW} & \colhead{$F$} & \colhead{EW} & \colhead{$F$} & \colhead{EW} &
\colhead{$F$} & \colhead{EW} } 
\startdata
0043 &  1.0 & 23 & 1.8 & 91 &\nodata&\nodata&\nodata&\nodata\nl
0748 &  1.7 & 13 & 1.7 & 27 & 0.12 & 1.0 & 0.5 & 4 \nl
0808 & 5.9 & 1.0 &\nodata  &\nodata& 8.4 & 1.0 &\nodata&\nodata\nl
0900 & 6.9 & 61 & 10 & 75 & 1.1 & 10 &\nodata&\nodata\nl
0904+03 & 0.2 & 3.0 & 0.7 & 26  &\nodata&\nodata&\nodata&\nodata\nl
0904+44 & 1.4 & 23 & 2.9 & 93 &\nodata&\nodata&\nodata&\nodata\nl
0932 & 8.9 & 43 & 9.2 & 74 & 5.3 & 23 & 5.8 & 26 \nl
0943 & 8.9 & 60 & 7.8 & 87 & 2.3 & 15 & 0.6 & 4 \nl
0951 & 1.2 & 46 & 1.6 & 89 &\nodata&\nodata&\nodata&\nodata\nl
1013 & 5.4 & 96 & 7.0 & 176 & 0.8 & 12 &\nodata&\nodata\nl
1020 & 16 & 2 & 19 & 5 &\nodata&\nodata&\nodata&\nodata\nl
1132 & 9.9 & 70 & 6.7 & 71 & 2.0 & 13 & 1.3 & 10 \nl
1152 & 7.4 & 46 & 11.3 & 127 & 0.4 & 2 &\nodata&\nodata\nl
1156 & 0.5 & 9 & 0.6 & 12 &\nodata&\nodata&\nodata&\nodata\nl
1216 & 3.6 & 73 & 7.0 & 245 &\nodata&\nodata&\nodata&\nodata\nl
1244 & 14.9 & 64 & 12.4 & 92 & 2.4 & 9 & 1.4 & 6 \nl
1249 & 2.5 & 3 & 4.3 & 8 &\nodata&\nodata&\nodata&\nodata\nl
1302 & 2.5 & 104 & 8.9 & 153 & 0.3 & 18 &\nodata&\nodata\nl
1305 & 2.1 & 38 & 2.0 & 51 & 0.5 & 8 & 0.1 & 2 \nl
1335 & 0.6 & 2 & 0.9 & 8 &\nodata&\nodata&\nodata&\nodata\nl
1422 & 4.6 & 97 & 3.5 & 73 & 1.0 & 19 & 3.9 & 63 \nl
1501 & 1.6 & 33 & 2.9 & 101 &\nodata&\nodata&\nodata&\nodata\nl
1538 & 13.5 & 84 & 12.7 & 167 & 2.4 & 12 & 0.8 & 5 \nl
1550 & 9.5 & 49 & 9.3 & 79 & 3.5 & 19 & 1.6 & 9 \nl
1610 & 2.0 & 53 & 5.5 & 320 &\nodata&\nodata&\nodata&\nodata\nl
1622 & 4.8 & 57 & 6.3 & 73 & 1.8 & 17 & 0.6 & 6 \nl
1626 & 2 & 11 & 1.9 & 22 &\nodata&\nodata& 1.9 & 10 \nl
1702 & 6.4 & 33 & 13.1 & 88 &\nodata&\nodata&\nodata&\nodata\nl
1703 & 11.6 & 56 & 14.7 & 73 & 3.1 & 17 & 1.3 & 7 \nl
1705 & 3.8 & 83 & 4.4 & 90 & 0.4 & 7 &\nodata&\nodata\nl
1711 & 1.3 & 31 & 2.4 & 97 &\nodata&\nodata&\nodata&\nodata\nl
2047 & 3.5 & 102 & 3.6 & 173 & 1.1 & 30 & 0.2 & 6 \nl
2100 & 0.4 & 2 & 0.9 & 8 &\nodata&\nodata&\nodata&\nodata\nl
2102 &\nodata&\nodata& 1.2 & 4 &\nodata&\nodata&\nodata&\nodata\nl
2116 & 0.4 & 38 & 0.8 & 143 &\nodata&\nodata&\nodata&\nodata\nl
2232 & 0.7 & 187 & 0.9 & 167 &\nodata&\nodata&\nodata&\nodata\nl 
\enddata
\tablenotetext{a}{Fluxes are in units of 10$^{-15}$ ergs cm$^{-2}$ s$^{-1}$,
equivalent widths are in units of \AA}
\end{deluxetable}

\subsection{Previously Known Systems}

As SDSS makes no attempt to screen out previously known sources in the
selection for objects that will receive fibers, known CVs are observed as well
as new systems. In 2002, the spectra of the known dwarf novae BC UMa (SDSS~J1152) and KS UMa (SDSS~J1020), 
as well at that of the 2dF source (SDSS~J1422) = 2QZ J142256.3-022108 
(Croom et al. 2001)
were obtained. Information on all three sources is 
available from the on-line CV catalog of Downes 
(http://icarus.stsci.edu/$\sim$downes/cvcat/). 
The detailed characteristics of SDSS~J1324 as an unusual low accretion 
rate
Polar have already been published by Szkody et al. (2003a). 

The SDSS spectrum of KS UMa was obtained during an outburst so it shows a
 typical disk-dominated spectrum. A spectrum of KS UMa at quiescence is shown 
by Jiang et al. (2000). 
 BC UMa shows the broad absorption features surrounding the emission
that is indicative of the prominence of the white dwarf (as previously noted by
Mukai et al. 1990). SDSS~J1422, on the other hand, has prominent \ion{He}{2}
emission and is
a likely candidate for a system containing a magnetic white dwarf.

\subsection{High Inclination Systems}

Generally, high inclination CVs have very prominent central
absorption in the Balmer lines (giving a double-peaked appearance),
 with increasing absorption up the Balmer series.
Figure 1 shows that SDSS~J0904, 1501 and 1702, are good candidates for high inclination,
eclipsing systems. Follow-up data on these 3 systems have confirmed that all
3 eclipse.  SDSS~J0904+03 shows a partial eclipse with a period of 86 min
and will be discussed in detail in a separate paper (Woudt et al. 2004).

\begin{figure}[!htb]
\resizebox{.48\textwidth}{!}{\rotatebox{0}{\plotone{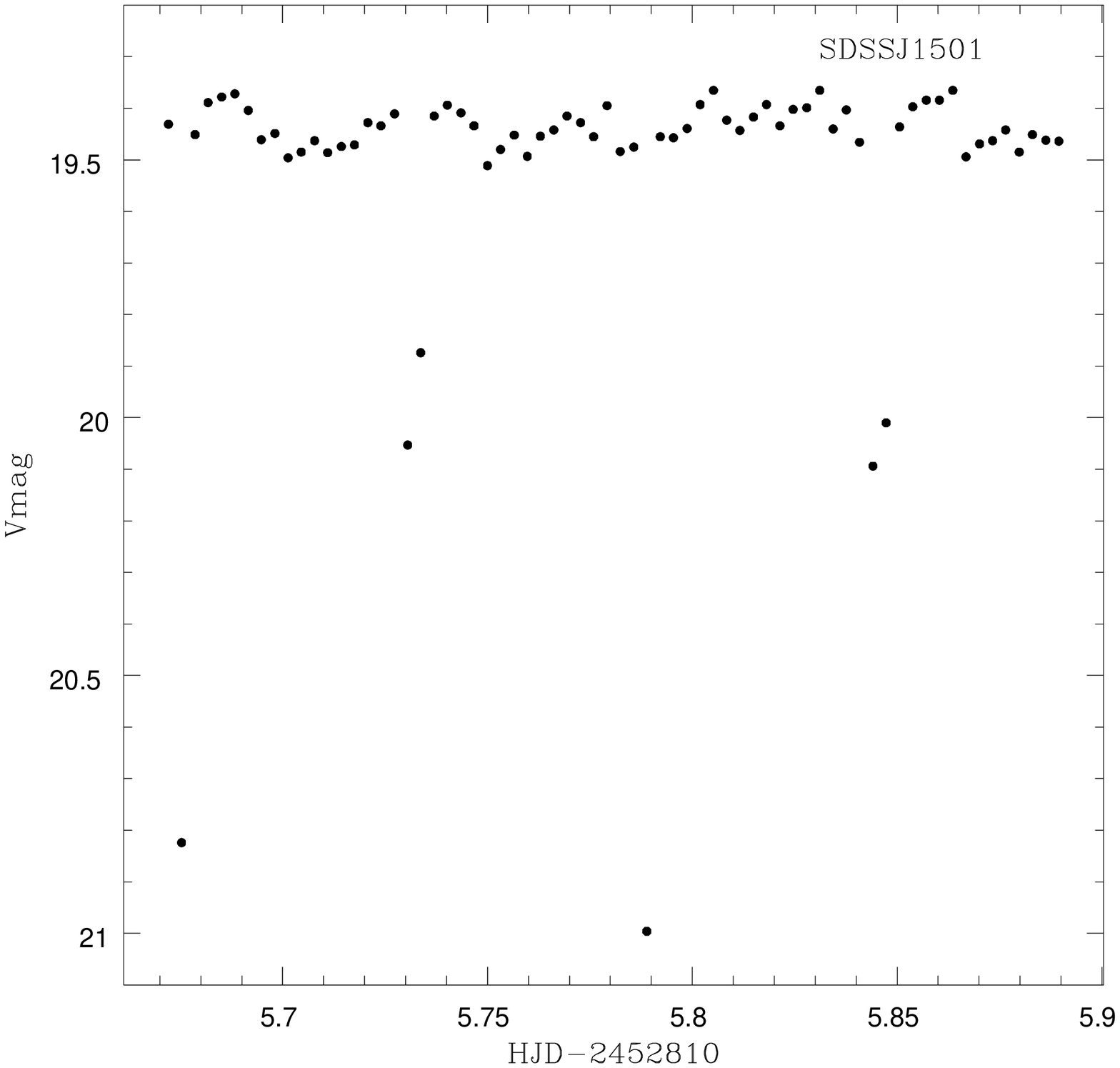}}}
\caption{NOFS light curve of SDSS~J1501. Integration times are 4 min and
photometric errors are 0.04 mag on each point except during eclipses}
\end{figure}

Photometry of SDSS~J1501 (Figure 2) reveals deep ($>$1.6 mag) 
eclipses with a period of 
78 min. Note from Figure 2, that the 4 min sampling does not resolve the
eclipses very well since they are of very short duration ($<$15 min).
The light curve is relatively flat between eclipses, indicating the lack of a 
prominent hot
spot where the mass transfer stream impacts the disk.

\begin{deluxetable*}{lcccccc}
\tablewidth{0pt}
\tablecaption{Radial Velocity Solutions}
\tablehead{
\colhead{SDSS~J} & \colhead{Line} & \colhead{$P$} & 
\colhead{$\gamma$} & \colhead{$K$} &
\colhead{$T_{0}$ } & \colhead{$\sigma$}\\
&&\colhead{(hr)}&\colhead{(km s$^{-1}$)}&\colhead{(km s$^{-1}$)}&\colhead{(JD 2,452,000$+$)}&
}
\startdata
0043 & H$\alpha$ & 1.5 & \phs$17\pm1$  & \phn$68\pm16$ & 909.830 & 28 \nl
0043 & H$\beta$ & 1.2 & \phn$-0.7\pm1.3$ & \phn$90\pm12$ & 909.783 & 27 \nl
0748 & H$\alpha$ & 2.5 & $-32\pm1$ & $177\pm23$ & 909.965 & 42 \nl
0748 & H$\beta$ & 2.5 & $-82\pm1$ & $157\pm45$ & 909.958 & 85 \nl
0900 & H$\alpha$ & 4.7 & \phs$54\pm7$ & $168\pm14$ & 327.706 & 36 \nl
0900 & H$\beta$ & 5.3 & \phs$89\pm9$ & $140\pm14$ & 327.707 & 39 \nl
0900 & H$\gamma$ & 4.5 & \phs$77\pm2$ & $126\pm12$ & 327.707 & 38 \nl
0932 & H$\alpha$ & 1.7 & \phn$-2\pm2$ & $109\pm25$ & 278.752 & 61 \nl
0932 & H$\beta$ & 1.7 & $-99\pm5$ & $168\pm30$ & 278.755 & 76 \nl
0932 & H$\gamma$ & 1.6 & $-14\pm6$ & $169\pm33$ & 278.755 & 83 \nl
1538 & H$\alpha$ & 1.7 & $-19\pm1$ & $103\pm4$\phn & 451.744 & 9 \nl
1538 & H$\beta$ & 1.5 & \phs$25\pm1$ & $160\pm12$ & 451.743 & 24 \nl
1538 & H$\gamma$ & 1.6 & $-87\pm2$ & $106\pm6$\phn & 451.748 & 11 \nl
1702 & H$\alpha$ & 2.5 & $-26\pm4$ & $110\pm10$ & 451.855 & 23 \nl
1702 & H$\beta$ & 2.5 & \phs$51\pm5$ & $124\pm13$ & 451.856 & 31 \nl
\enddata
\end{deluxetable*}

\begin{figure}[!tb]
\resizebox{.46\textwidth}{!}{\rotatebox{0}{\plotone{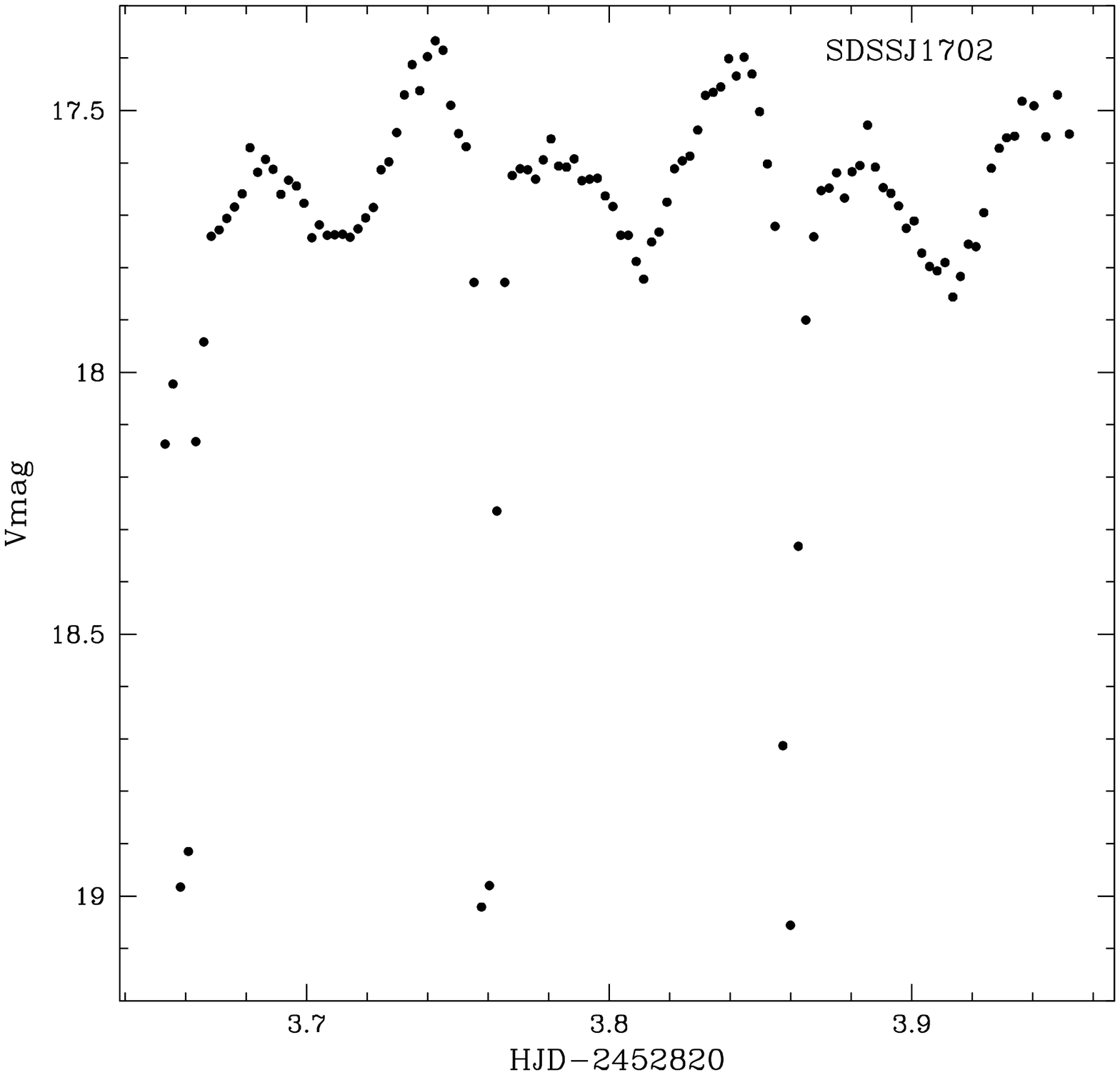}}}
\caption{NOFS light curve of SDSS~J1702. Integrations are 3 min with errors
of 0.01 mag outside of eclipse.}
\end{figure}

The light curve of SDSS~J1702 (Figure 3) is particularly interesting. It shows 
deep (1.5 mag) eclipses with a period of 2.5 hrs, a secondary eclipse of
0.2 mag depth at phase 0.5, and a large (0.2 mag) hump preceding eclipse
which is likely the hot spot coming into view. Using the double-Gaussian
method on the emission 
line wings and fixing the period at 2.5 hrs gives the radial
velocity solutions listed in Table 4. Figure 4 shows the best fits to the
H$\alpha$ and H$\beta$ lines. Note that the points near phase 0.9 (which
are not included in the fit) show the usual large displacement to
positive then to negative velocity that is
 typical for accretion disks undergoing
an eclipse. 
\begin{figure}[!tb]
\resizebox{.46\textwidth}{!}{\rotatebox{0}{\plotone{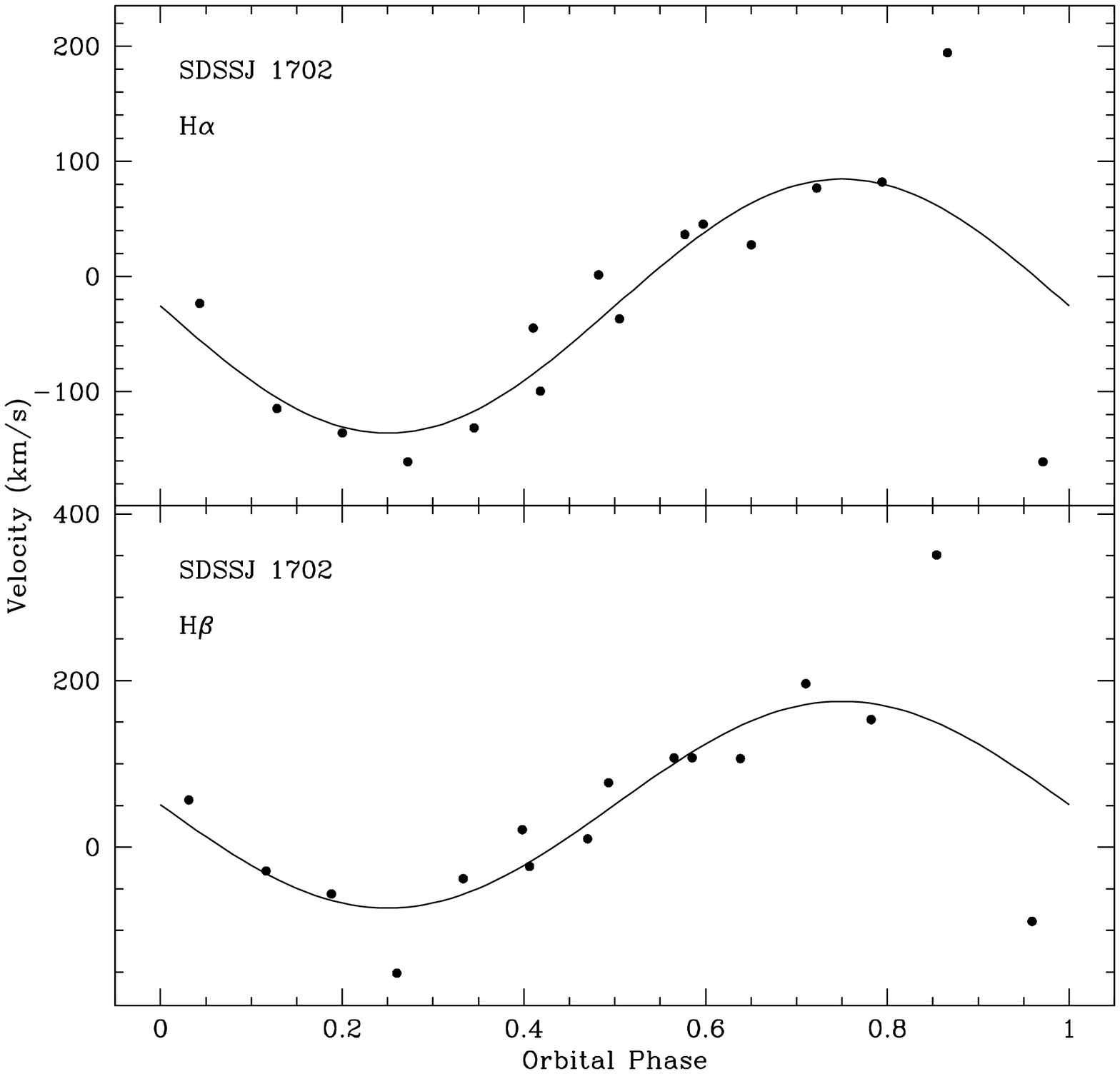}}}
\caption{Velocity curves of SDSS~J1702 with the best fit sinusoids (Table 4)
plotted on the data. The two discrepant points near phase one are not 
included in the sine fit.}
\end{figure}
SDSS~J1702 is an excellent candidate for
further follow-up observations with a large telescope; better time-resolution
for the light curve should be able to resolve the eclipse structure, which
may show
the white dwarf ingress and egress. The width of the eclipse combined with
the radial velocity curve may provide an estimate of the white dwarf mass.
In addition, Figure 1 shows that the M star is visible. Matching the 7050 TiO 
band to the spectral index defined by Reid, Hawley \& Gizis (1995)
for M stars yields a spectral type of M1.5$\pm$1.1. Then, using the absolute
magnitude vs spectral type relations from Hawley et al. (2002) with the
$\it{i}$ and $\it{z}$ magnitudes listed in Table 1 yields a distance of 610 pc from 
the $\it{i}$ magnitude, 580 pc from the $\it{z}$ magnitude, and 570 pc using the $\it{J}$ 
magnitude
from the 2MASS database ($\it{J}$=15.63$\pm$0.05). With the uncertainties in the
spectral type, the total range of distances allowed is between 460-650 pc. 

 An attempt was made to determine the velocity of the M dwarf 
by using the IRAF routine $\it{fxcor}$ to cross-correlate 
the red spectra against the one in eclipse, but the individual spectra were too
 noisy to derive a believable result. Time-resolved
higher S/N data should be able to determine the velocity curve of the
secondary, thus leading to a good determination of the mass ratio. 
SDSS~J1702 is potentially very useful, as it is one of the very
few double-lined, eclipsing CVs, from which masses can be found.
A mass measurement for this system would be
particularly interesting, as its period is
squarely in the period gap (a region between 2-3 hrs that is rarely occupied
by dwarf novae). SDSS~J1702 can provide a good test of the evolution models of Howell, Nelson \&
Rappaport (2001), which make predictions for the masses above and below the gap.

\subsection{Dwarf Novae}
\begin{figure}[!tb]
\resizebox{.46\textwidth}{!}{\rotatebox{0}{\plotone{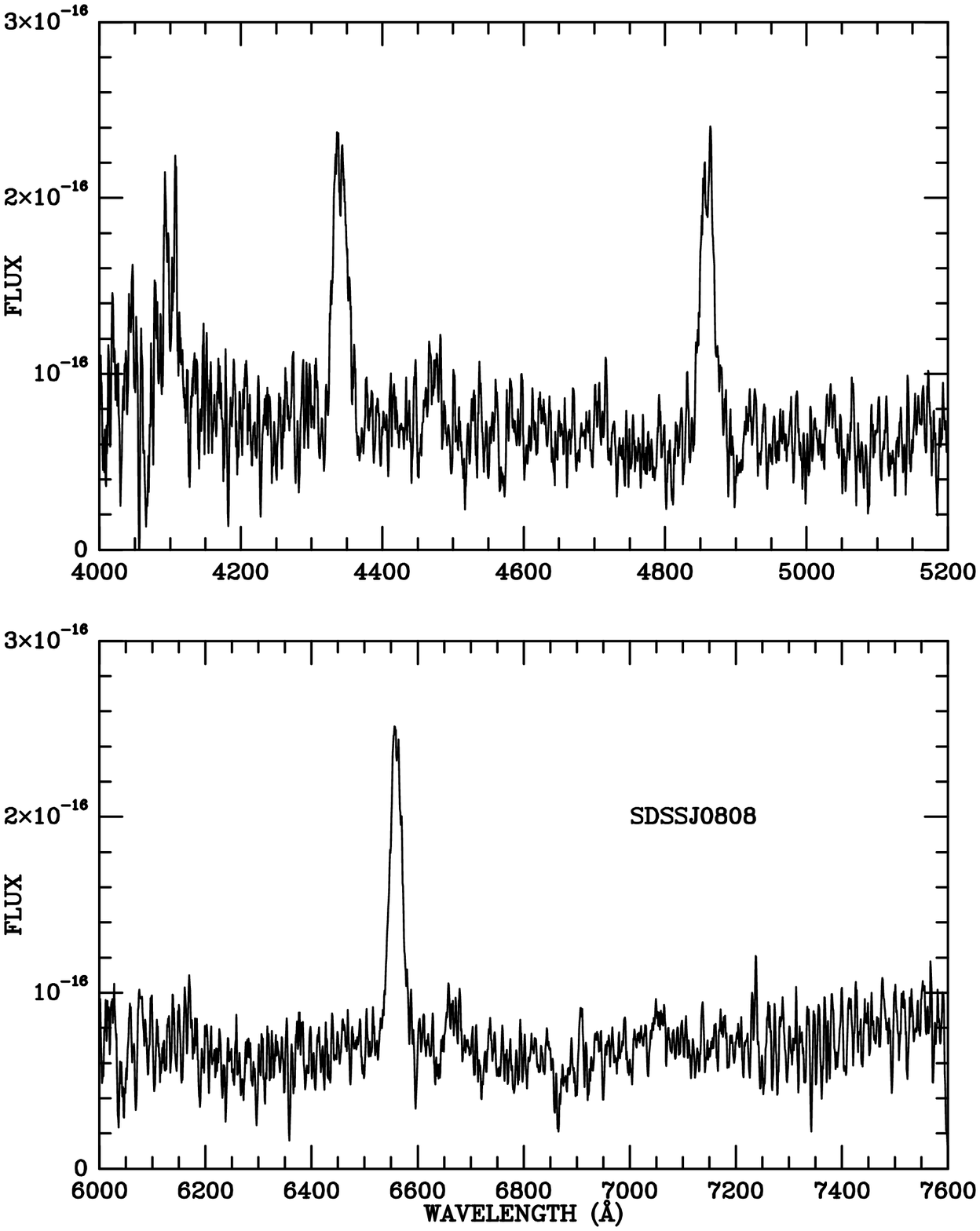}}}
\caption{APO quiescent spectra of SDSS~J0808 obtained 2003 Nov 27.}
\end{figure}

The SDSS spectrum of SDSS~J0808 was obtained during a likely 
dwarf nova outburst and
is similar to the SDSS outburst spectrum of KS UMa (SDSS~J1020). 
Both the SDSS image and the Digitized Sky Survey (DSS)
show SDSS~J0808 as a fainter object than implied by the spectrum in Figure 1.
An APO spectrum obtained on 2003 Nov 27 (Figure 5) shows the typical
strong emission lines of a quiescent dwarf nova. It is likely that this object
has fairly frequent oubursts, as follow-up photometry at USNO also caught SDSS~J0808
during outburst (at $\it{V}$=14.5). The outbursts are fairly long as an observation a
week after the first one showed the system still at 14.7. The 9 hours of
observation 
at that time (Figure 6)
shows a modulation with amplitude of 0.04 mag on a timescale of 6 hrs 
superimposed
on a declining brightness trend throughout the observation. However,
since it is usually not possible to determine an orbital period from
outburst photometry and a radial velocity curve is not yet available, 
futher data will
be needed to determine if the 6 hr timescale is related to the orbital 
period of this system,
or merely transient fluctuations in the mass accretion as the system returned
to quiescence.
\begin{figure}[!tb]
\resizebox{.46\textwidth}{!}{\rotatebox{0}{\plotone{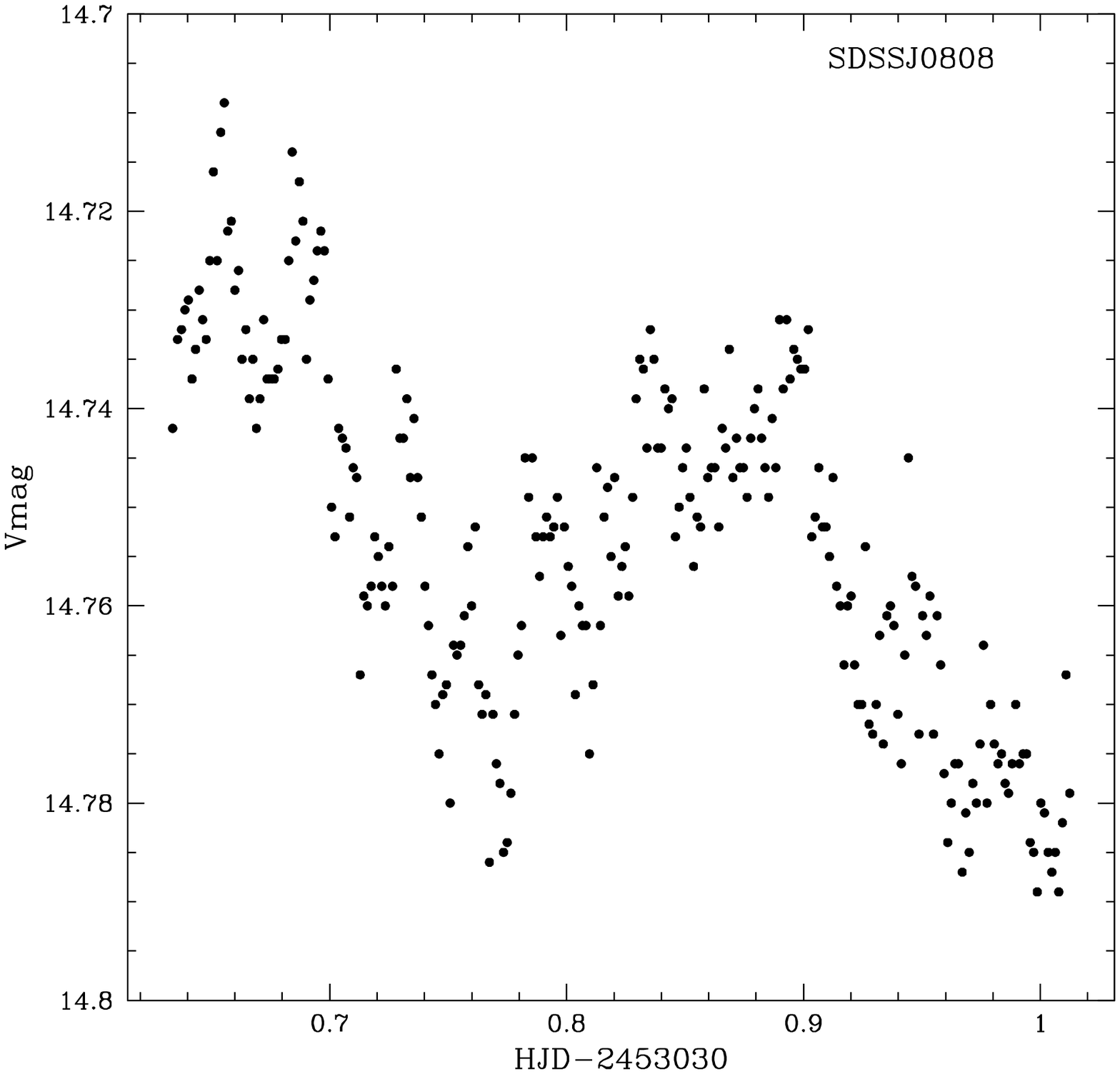}}}
\caption{NOFS light curve of SDSS~J0808 obtained near outburst in a V
filter. Integration times are 1.5 min and photometric errors are 0.004 mag.}
\end{figure}

The SDSS spectrum of SDSS~J2100 (Figure 1) also looks similar to an
outbursting dwarf nova and the SDSS image and photometry (Table 1)
 is fainter as well. Optical
photometry over several nights and during different accretion states 
revealed a large amplitude (0.2-0.5 mag) periodic
modulation at 
1.9 hrs (or a possible alias at 2.1 hrs); the detailed results are discussed in
a separate paper (Tramposch et al. 2004).

SDSS~J2232 was found in ouburst during the SDSS imaging (Table 1) while
it was in quiescence during the SDSS spectrum (Figure 1). USNO photometry
during quiescence on 2003 January 8 and 12 showed V=22.4, B=22.8$\pm$0.5 mag
while an outburst (V=18.45) was caught on 2003 Oct 20 (Table 2). The time 
series during outburst did not reveal any modulation.

Lastly, SDSS~J2102 has the appearance of a dwarf nova during outburst, but
with exceptionally strong Balmer absorption. Additional spectra obtained at APO
in 2003 December showed a similar appearance and the SDSS and DSS images
show comparable brightness. Thus, it is more likely that this is a system with a
prominent white dwarf. It is possible that this could be a detached system,
although H$\alpha$ is broader than a typical line caused by irradiation
of the secondary by the white dwarf. Time-resolved spectroscopy can reveal
if this is a long period, widely separated binary.

\subsection{Nova-likes with Strong \ion{He}{2}}

Strong \ion{He}{2} emission lines usually indicate either a 
magnetic white dwarf, where the 
EUV continuum from an accretion shock ionizes the plasma, 
or a novalike system with a high level of mass transfer, often a type of
system termed an SW Sex star. These types are ultimately differentiated by
polarimetry, although time-resolved photometry and spectroscopy can yield
distinct clues. An object that has 
a large circular polarization and a strong modulation of the lines and 
continuum at the orbital period is classified as a spin-orbit synchronized 
AM Her system, or Polar (see review of Polars in Wickramasinghe \& Ferrario
2000).  One that has little or no polarization and is observed to have a  
periodicity at a spin timescale of the white 
dwarf (minutes), is considered an Intermediate Polar (IP), while one with an 
orbital period between 3-4 hrs and single-peaked emission lines is
likely an SW Sex star (see Warner 1995 for a review of all types). 

The systems SDSS~J0748, 0932, 1422, and 1626 clearly show \ion{He}{2}.
Of these,
 SDSS~J1422 (the 2dF source) and SDSS~J1626
are the best candidates for Polars, as they have \ion{He}{2} greater than H$\beta$. 
Spectropolarimetry was accomplished for
SDSS~J0748 but no polarization was seen to a level of 0.04\%, so a Polar nature is
ruled out for this object. Spectroscopy suggests that the orbital period could
be on the
order of 2.5 hrs (Figure 7 and Table 4), but the dataset is only 2.6 hrs long so
further spectroscopy is needed for a correct determination. 
\begin{figure}[!tb]
\resizebox{.46\textwidth}{!}{\rotatebox{0}{\plotone{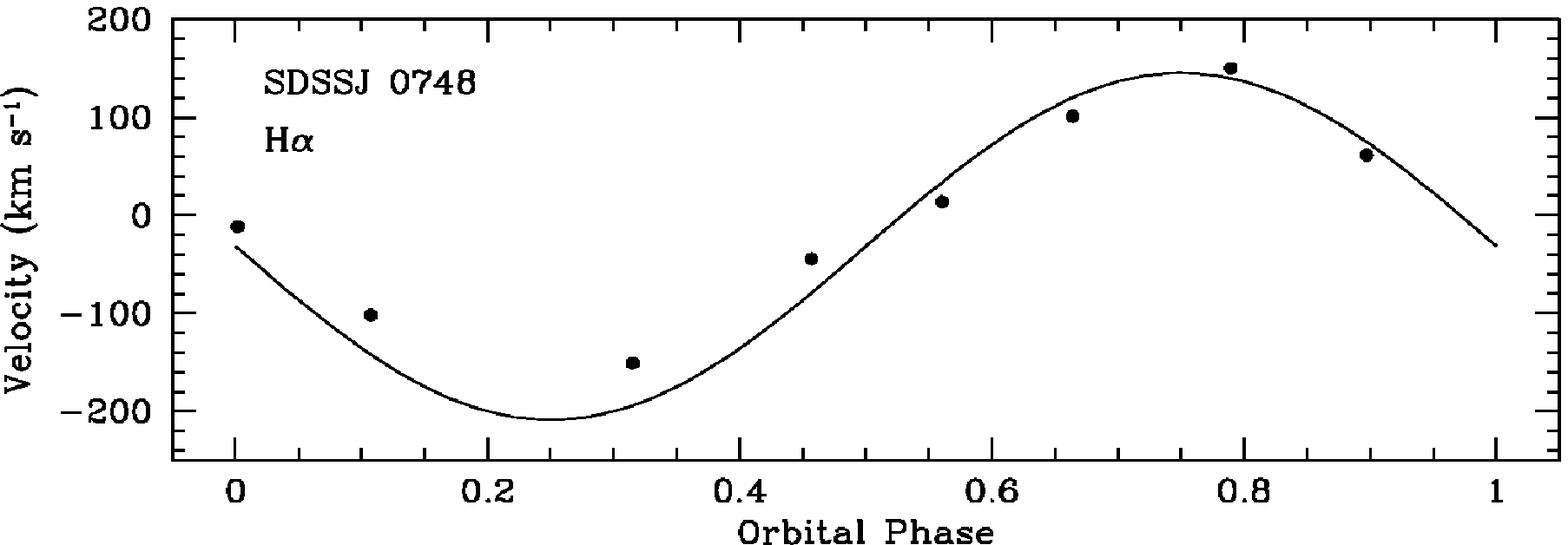}}}
\caption{H$\alpha$ velocity curve of SDSS~J0748 with the best fit sinusoid
from Table 4.}
\end{figure}
While \ion{He}{2} emission was present in
the APO spectra, the poor S/N due to high airmass at the time of observations 
prevented any determination of the velocities or period from the lines in the 
blue portion of the spectrum. Due to the proximity of a close companion on
the sky, the photometry of this object was accomplished with an aperture
that included both stars.\begin{figure}[!tb]
\resizebox{.46\textwidth}{!}{\rotatebox{0}{\plotone{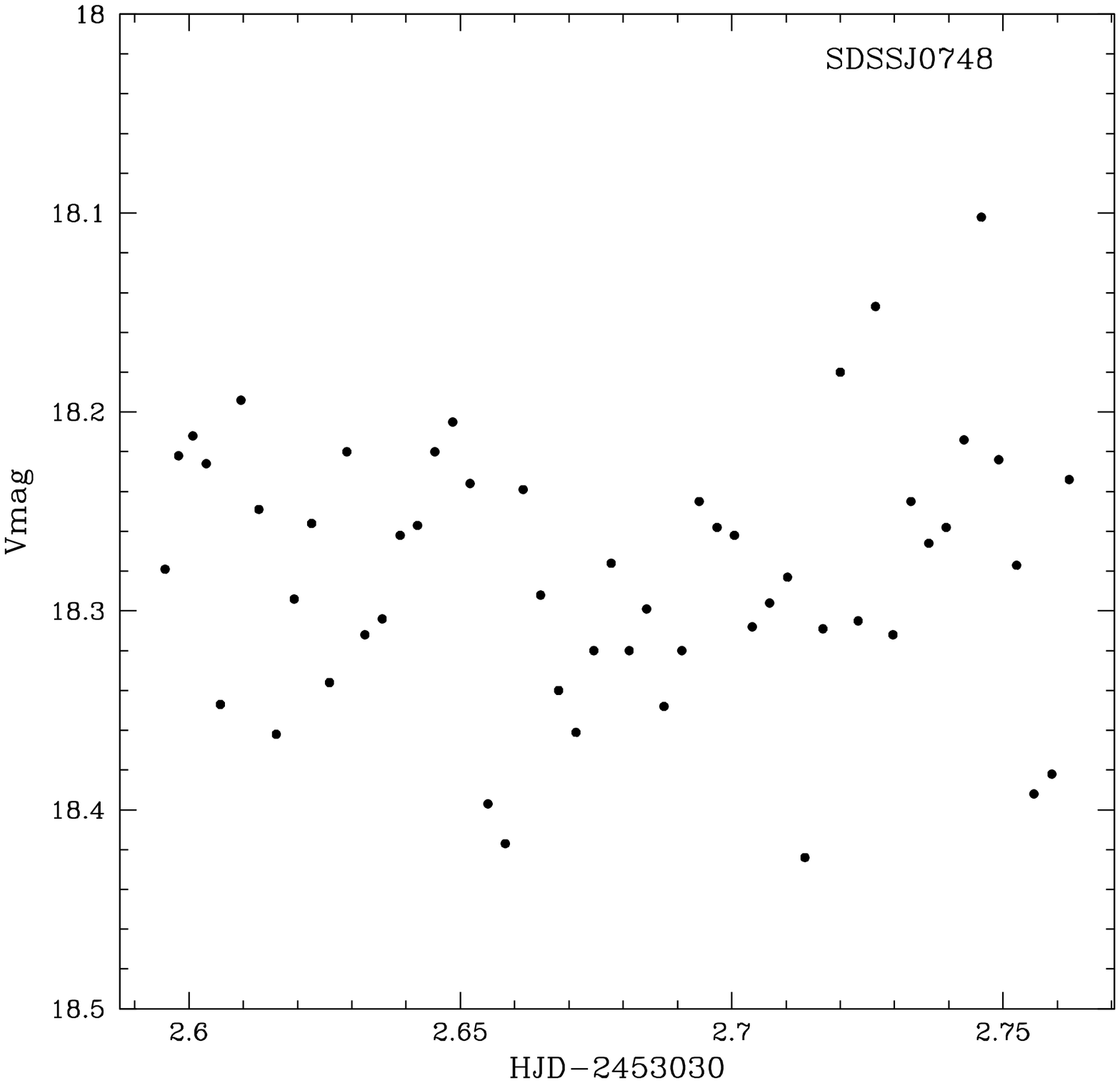}}}
\caption{NOFS light curve of SDSS~J0748. Each point is a 3 min integration
with error bar of 0.03. The aperture used included both the object and
a close (fainter) companion on the sky.}
\end{figure}
 The light curve (Figure 8) shows no clear
modulation at the 2.5 hr period. However, the $\it{pdm}$ (epoch folding)
period search method available in IRAF reveals a
prominent minimum at a period of 70 min. Since the light curve is only 4 hrs 
long,
further photometric data is also
 needed to determine if this is merely active flickering
or a stable period that could be 
identified with the spin of the white dwarf (which would indicate this
system could be an IP). 

\begin{figure}[!tb]
\resizebox{.46\textwidth}{!}{\rotatebox{0}{\plotone{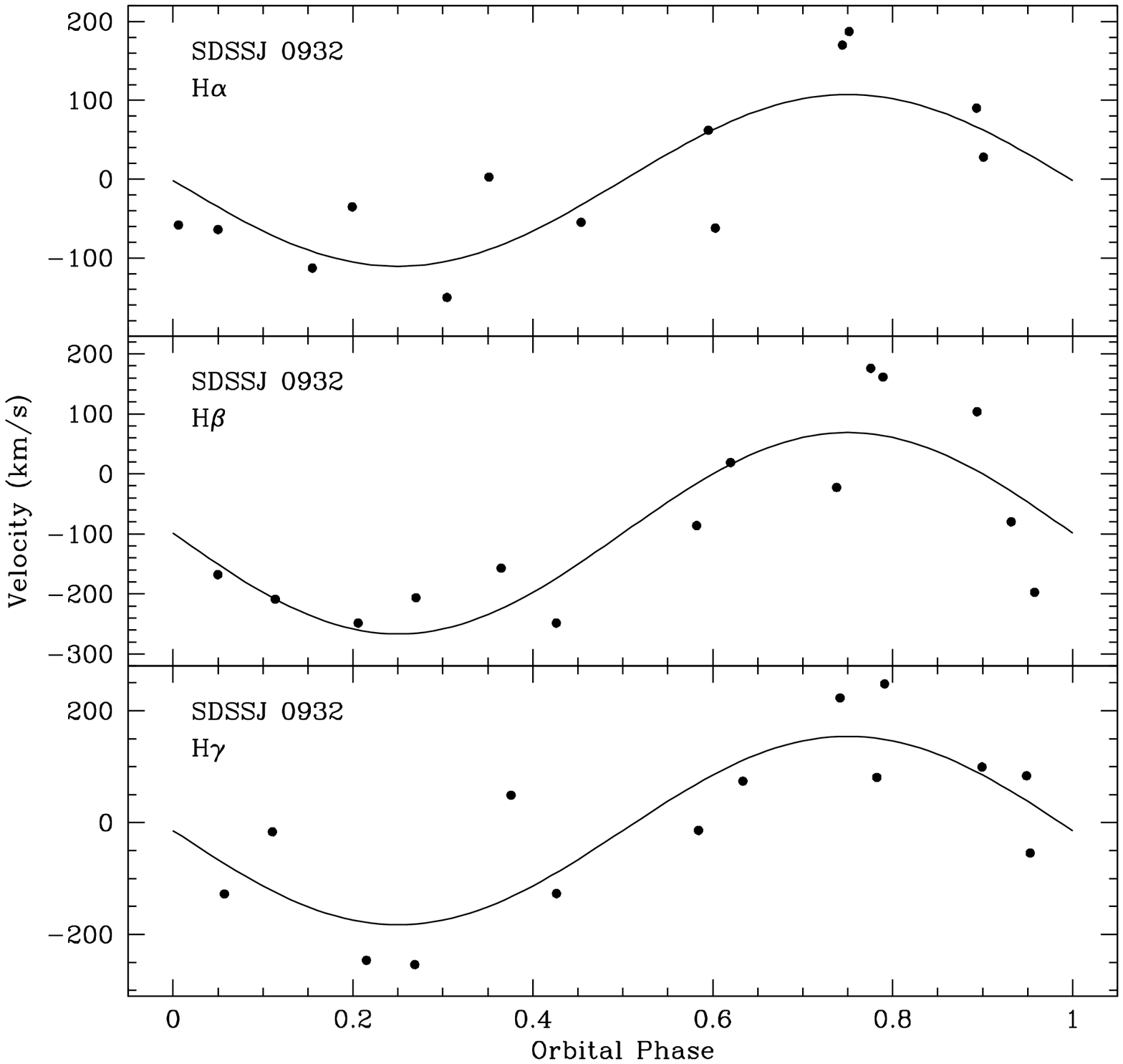}}}
\caption{Velocity curves of SDSS~J0932 with the best fit sinusoids from
Table 4.}
\end{figure}
{\bf{\it SDSS~J0932}}: This object was discovered as US 691 in the Usher survey of
 faint blue objects at high galactic latitude
 (Usher, Mattson \& Warnock 1982). Its spectrum is very unusual in that it
 shows very strong doubled helium emission lines  
as well as the Balmer hydrogen lines. This doubling implies an origin in a disk.
Neither Polars nor SW Sex stars show this behavior. The unusual strength and
structure in the He lines in SDSS~J0932 is reminiscent
of the SDSS-discovered CV SDSS~J102347.67+003841.2 (Bond et al. 2002; Szkody et 
al. 2003b). It is not clear why these 2 systems have such strong and
doubled He lines in comparison to normal novalikes.
 While SDSS~J1023 is a FIRST radio source, implying a magnetic nature,
there is no radio source within 30 arcsec of SDSS~J0932. However, observations
have shown that these radio sources can be highly variable (P. Mason, 2004, 
private
communication).  Our radial velocity
data (Figure 9 and Table 4) give a best fit to a sine curve suggesting a period 
near 1.7 hrs, but the overall errors are large for the brightness level of
this system, so further data are definitely needed.
It would be useful to obtain a longer set of time-resolved spectra with
higher time resolution in order to construct tomograms that could help to
locate the \ion{He}{2} emission. Usually, this line is produced close to
the white dwarf and provides the best velocity measurement 
value for the white dwarf. If the inner disk is disrupted, the line may
originate further out in the disk (giving a doubled appearance), but must
still be close enough to receive the excitation provided by high energy
photons. 

\subsection{Systems Showing the Underlying Stars}

Generally, the systems with the lowest mass transfer rates have accretion disks
that are so tenuous that they contribute little light in comparison to the
primary and secondary stars.  
 In these systems, the broad Balmer absorption lines from the 
white dwarf 
are observed flanking the
emission lines from the disk. TiO bands from the secondary star can also be 
present, as well as an upturn in the flux longward of 7000\AA. In the SDSS 
spectra obtained 
during 2002, there are many more such systems present than in the previous 2 
years of data.
The 15 systems showing one or both of the underlying stars include
SDSS~J0043, 0900, 0904+03, 0904+44, 0951, 1152 (BC UMa), 1249, 
1302, 1324, 1501, 
1610,
1702, 1705, 1711 and 2232. Three of these systems are the eclipsing objects
discussed in section 3.2, since a
disk viewed 
edge-on (i.e. high inclination) is not as luminous as a disk seen face-on
(Warner 1987).
It is notable that several of the white dwarfs in the 
systems showing prominent
absorption lines have been shown to be pulsating during  
follow-up studies conducted by Woudt \& Warner (2004) and  Warner \& Woudt (2004).
\begin{figure}[!tb]
\resizebox{.46\textwidth}{!}{\rotatebox{0}{\plotone{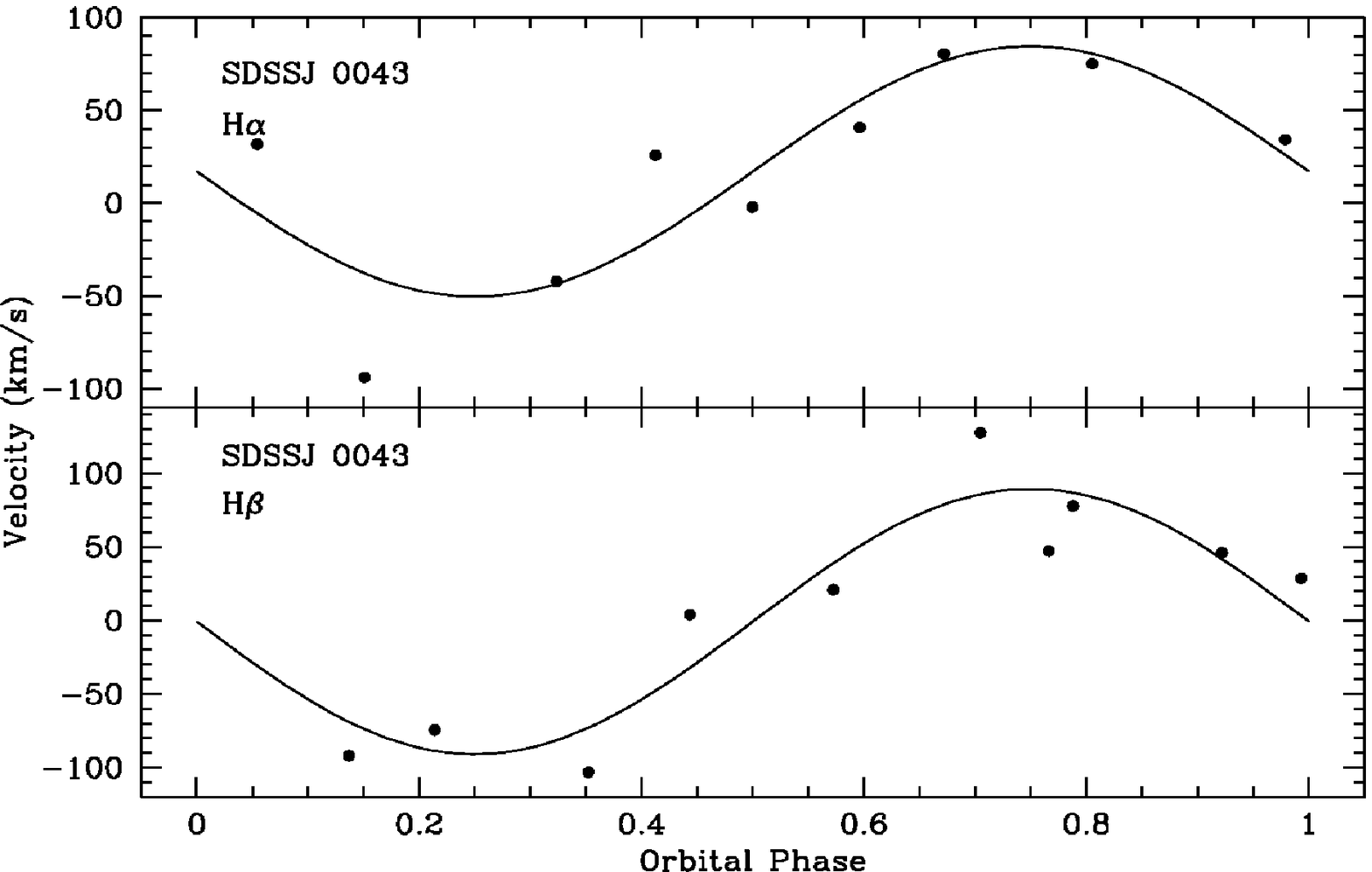}}}
\caption{Velocity curves of SDSS~J0043 with the best fit sinusoids from
Table 4.}
\end{figure}

{\bf{\it SDSS~J0043}}: Using the emission lines of H$\alpha$ and H$\beta$, our
2.7 hrs of data suggest a 
period near 1.5 hrs from H$\alpha$ and near 1.2 hrs from H$\beta$ (Table
4). Since H$\alpha$ is a stronger line and is not affected by the surrounding
absorption of the higher Balmer series, it represents the best solution,
but a longer time series of observations is needed to accurately pin down
the period. Figure
 10 shows the best-fit radial velocity curves for both lines.
 
\begin{figure}[!tb]
\resizebox{.46\textwidth}{!}{\rotatebox{0}{\plotone{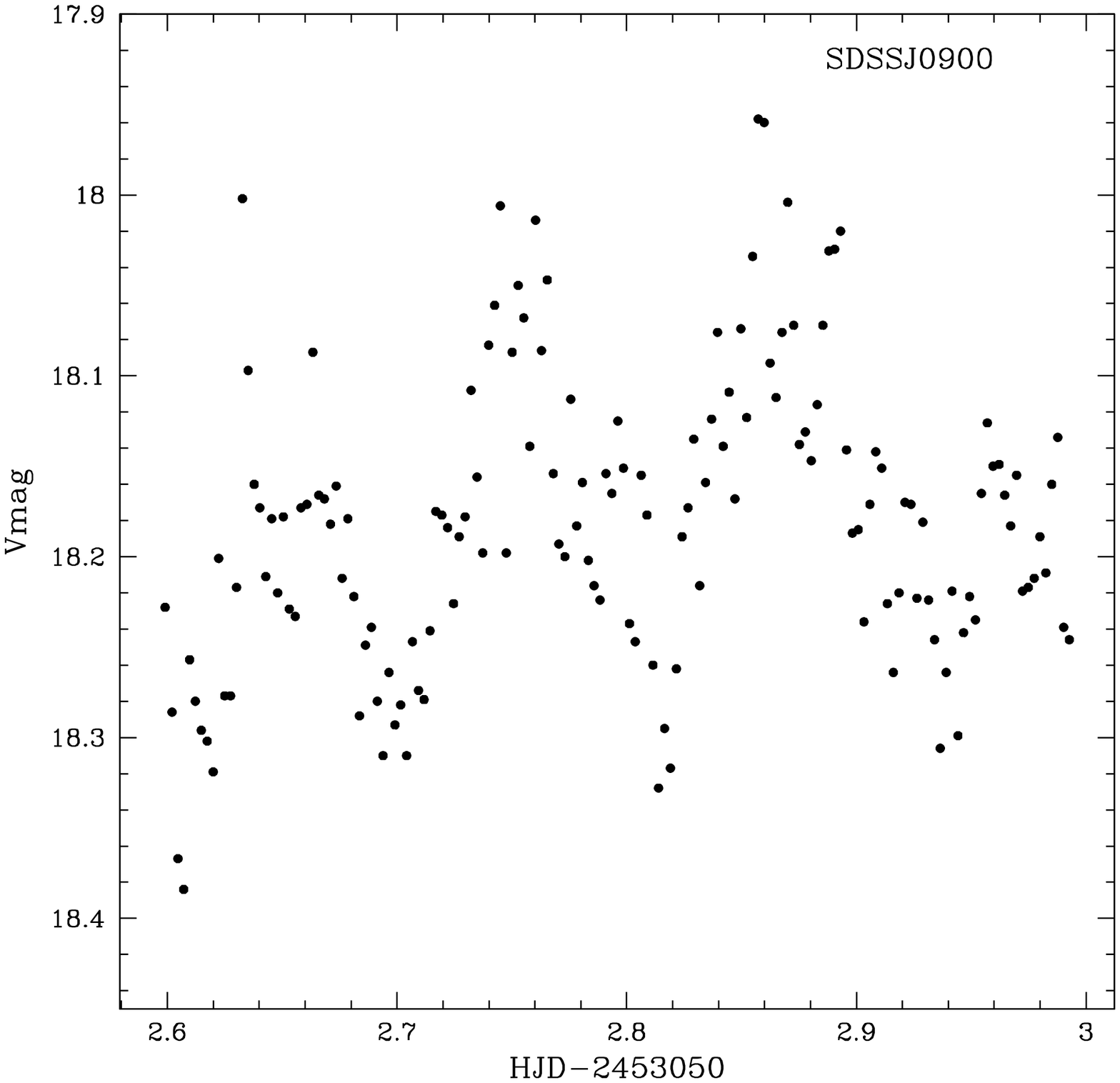}}}
\caption{NOFS light curve of SDSS~J0900 showing the 2.6 hr modulation. 
Integration times are 3 min with 0.03 mag errors.}
\end{figure}
{\bf{\it SDSS~J0900}}: This object is unusual in that it shows a prominent M star but
does not show any evidence for the white dwarf. Either the white dwarf is
very cool, the M star is evolved, or the accretion rate is relatively high 
in this system so that it obscures the white dwarf. The light curve shows a
modulation of 2.6 hrs (Figure 11), whereas the radial velocity solution
(Table 4 and Figure 12) suggests the period is twice this value. 
\begin{figure}[!tb]
\resizebox{.46\textwidth}{!}{\rotatebox{0}{\plotone{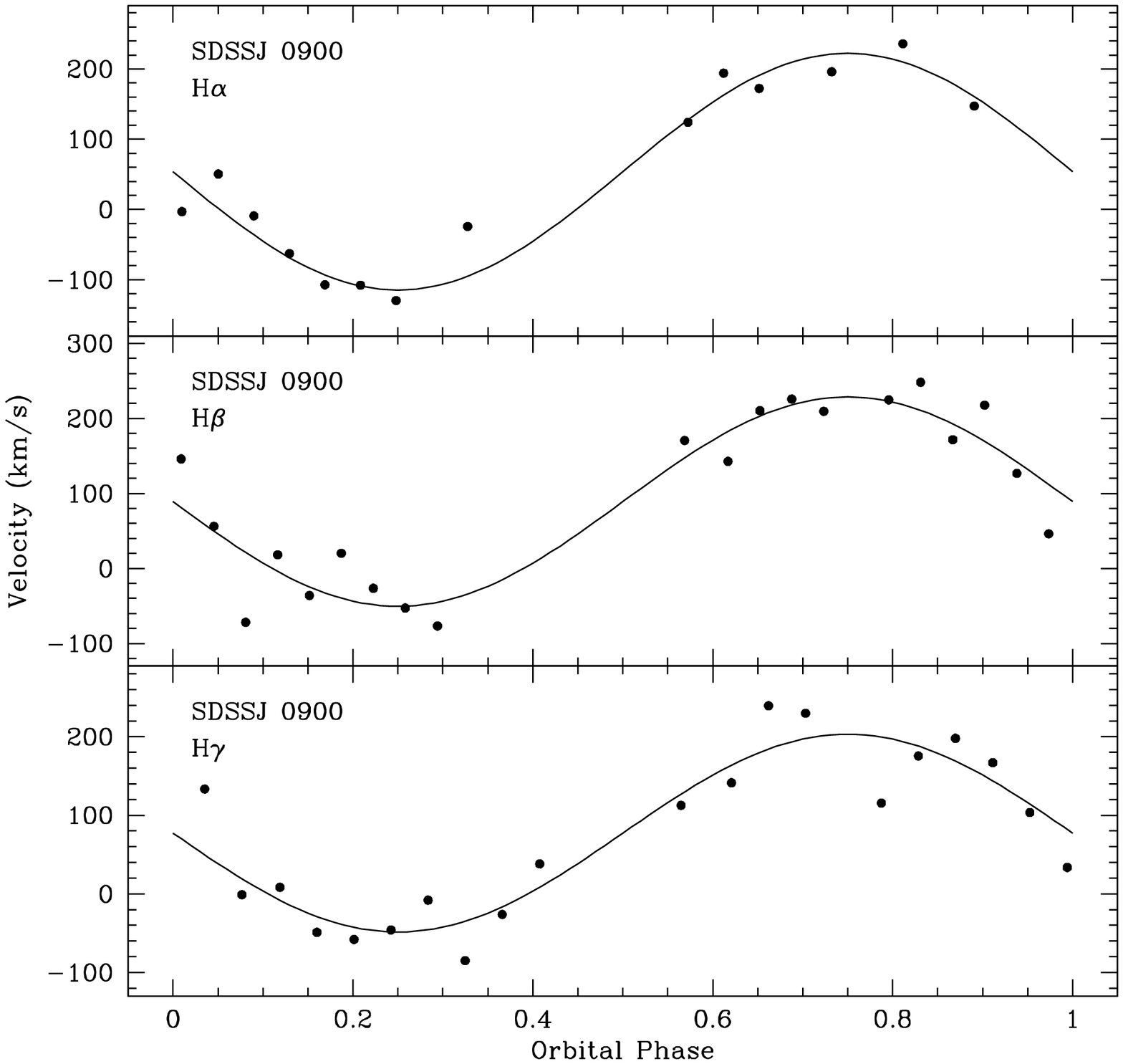}}}
\caption{Radial velocity curves of SDSS~J0900 with the best fit sinusoids
from Table 4. Note the period is twice that of the photometry.}
\end{figure}
The most logical interpretation is that the light curve is showing the
ellipsoidal variation due to the prominent secondary. 

{\bf{\it SDSS~J1249}}: While this object is listed in the catalog of Kilkenny,
Heber \& Drilling (1988) as a hot sub-dwarf, the SDSS spectrum clearly
shows the presence of an M star. The Balmer lines all have narrow, strong
emission line cores. Thus, this system appears to have a
hot star in close proximity to a highly irradiated M star. While there may
not be mass transfer going on, this system may be a pre-CV or a system
undergoing a time of low mass transfer.

{\bf{\it SDSS~J1302}}: This system's spectrum resembles that of a symbiotic star (not strictly a
CV), with a prominent M companion and very strong narrow emission lines
with a recombination decrement. However, this object lacks the high excitation
lines (e.g. \ion{He}{2}) typical of symbiotics (see reviews of symbiotic stars
by Corradi, Mikolajewska \& Mahoney 2003).

\subsection{Undetermined Objects}

The object SDSS~J1156 defies ready classification. While it has strong
Balmer emission, the lines are very narrow for a CV and the continuum
turns over in the blue. It also appears to have an emission line of \ion{He}{1}
$\lambda$5876 that is unusually strong. Since this object is quite faint, it could
be an accreting system in a low state, such as a Polar. Further observations
will be needed to sort out its correct identification.

APO spectra suggest a period near 1.6 hr for SDSS~J1538 (Table 4 and Figure 13).
\begin{figure}[!tb]
\resizebox{.46\textwidth}{!}{\rotatebox{0}{\plotone{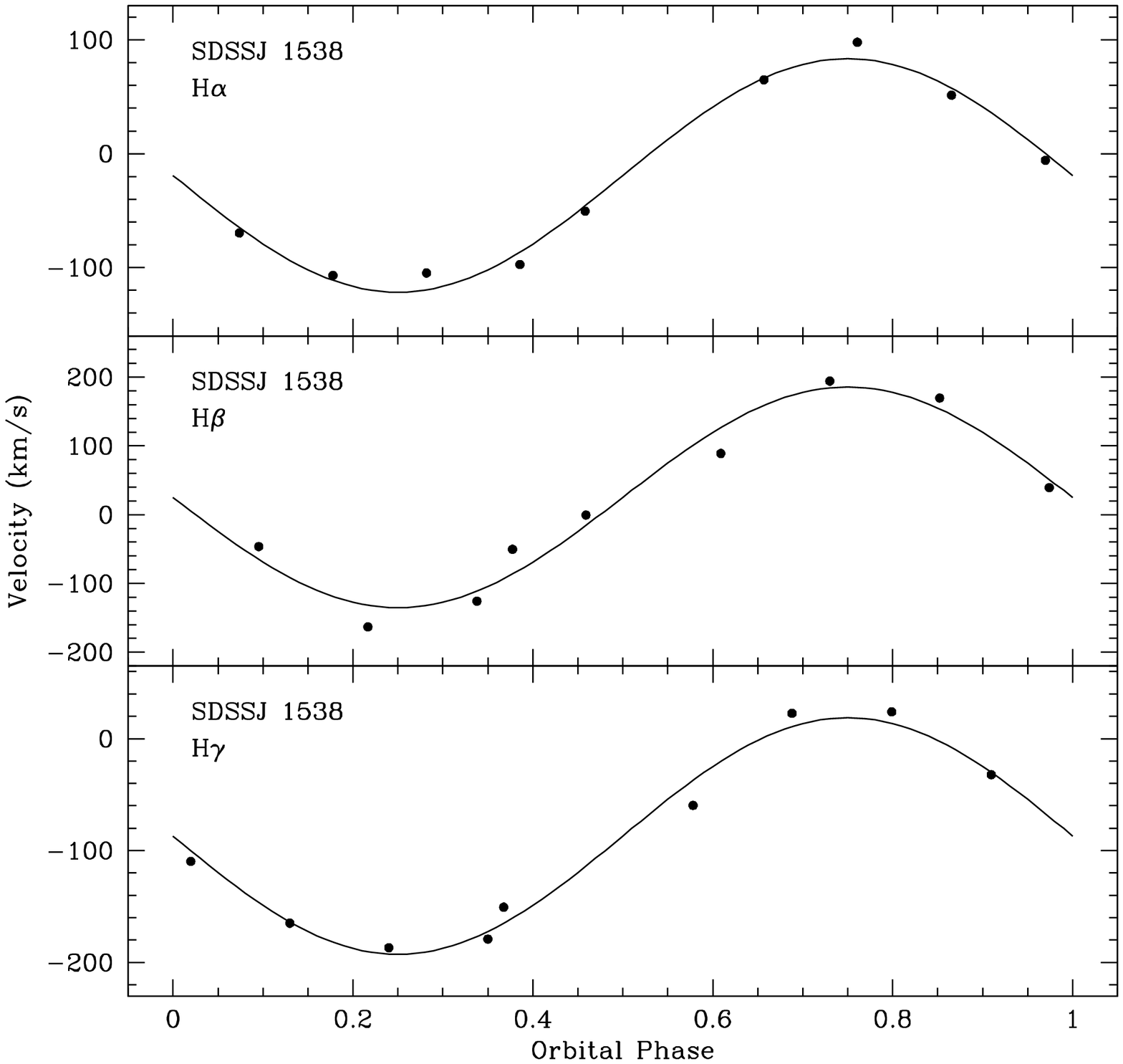}}}
\caption{Radial velocity curves of SDSS~J1538 with the best fit sinusoids
from Table 4.}
\end{figure}
The spectral appearance (Figure 1) and short period are typical for
a normal dwarf nova. Long term photometric monitoring can reveal if this
system shows typical dwarf nova outbursts and monitoring during an outburst
can reveal if it shows the superhump phenomena typical of those systems
with periods less than 2 hrs (SU UMa systems).
 
\subsection{ROSAT Correlations}
\begin{deluxetable}{lccl}
\tablewidth{0pt}
\tablecaption{ROSAT Detections}
\tablehead{
\colhead{SDSS~J} & \colhead{ROSAT} & \colhead{Exp}
& \colhead{1RXS J}\\
&  \colhead{ (counts s$^{-1}$)\tablenotemark{a}}&\colhead{(s)}&}
\startdata
0943 & $0.03\pm0.01$ & 544 & 094327.7+520139 \nl
1020 & $0.12\pm0.02$ & 560 & 102027.1+530439 = KS UMa \nl
1132 & $0.025\pm0.009$ & 576 & 113213.3+624903 \nl
1244 & $0.03\pm0.01$ & 533 & 124424.5+613511 \nl
1538 & $0.07\pm0.01$ & 735 & 153818.5+512348 \nl
\enddata
\tablenotetext{a}{For a 2 keV bremsstrahlung spectrum, 1 c/s corresponds to a
0.1-2.4 keV flux of about 7$\times10^{-12}$ ergs cm$^{-2}$ s$^{-1}$}
\end{deluxetable}

Matching the coordinates of the CVs in Table 1 with the X-ray 
ROSAT All Sky Survey
(RASS; Voges et al. 1999, 2000) reveals that 5 have
X-ray detections ($>$2.5$\sigma$) within the positional errors of the RASS. 
Table 5 presents the X-ray count rates for the 5 sources.
The previously known dwarf nova KS UMa is included in this list.
The low flux levels are typical for faint CVs with accretion disks. 
Surprisingly,
none of the candidates for magnetic systems, which usually have larger
X-ray fluxes than disk systems, appear in this list.

\section{Conclusions}

The 36 CVs described here, combined with the 64 from Papers I and II provide
substantial additions to the database of low luminosity, short orbital
period CVs that are postulated to comprise the true population of CVs
(Howell, Nelson \& Rappaport 2001). While the overall detection rate (0.03/square
degree) is comparable to that in Papers I and II, this latest set of CVs
 has fewer magnetic
candidates (4) but many more systems (15)
 showing the underlying stars than in the
past papers. Of the 12
systems with periods estimated (listed in Table 1), 8 are less than 2 hrs, 3 are in the period
gap between 2-3 hrs and only one is above 3 hrs. If the period estimates
are confirmed and are representative of the rest of the group,
the proportion of systems with
short  periods in the SDSS data will be larger than 
that from brighter surveys such as the Palomar-Green (Green et al. 1982;
online Downes CV catalog gives periods) or
the recent Hamburg Quasar Survey
(G\"ansicke, Hagen \& Engels 2002). This would be consistent
with the population models that predict most CVs should have evolved to
short orbital periods and low mass transfer rates. 

This sample highlights several interesting candidates for further study.
Of the 3 eclipsing systems
found, SDSS~J1702 is especially important in that the secondary M1.5 star
is evident so that the masses may be determined with a large telescope.
Further photometry of SDSS~J0748 is needed to determine if it could be an IP.
The exact nature of systems such as SDSS~J1156, 1335 and 2102 remain to
be explored. 
 In
addition, fitting of the underlying stars in the 15 systems that show them
 can reveal the
temperatures and distances of this sample, yielding some estimate of age
and heating effects from the ongoing accretion. 

\acknowledgments

Funding for the creation and distribution of the SDSS Archive has been provided 
by the Alfred P. Sloan Foundation, the Participating Institutions,
the National Aeronautics and Space Administration, the National Science 
Foundation, the U.S. Department of Energy, the Japanese
Monbukagakusho, and the Max Planck Society. The SDSS Web site is 
http://www.sdss.org/.
The SDSS is managed by the Astrophysical Research Consortium (ARC) for the 
Participating Institutions. The Participating Institutions are The
University of Chicago, Fermilab, the Institute for Advanced Study, the 
Japan Participation Group, The Johns Hopkins University, Los Alamos
National Laboratory, the Max-Planck-Institute for Astronomy (MPIA), the 
Max-Planck-Institute for Astrophysics (MPA), New Mexico State
University, University of Pittsburgh, Princeton University, the US 
Naval Observatory, and the 
University of Washington. 
P.S., N.S. and S.H. acknowledge support from NSF grant AST 02-05875.
Studies of magnetic stars and stellar systems at Steward Observatory is supported by the NSF through AST 97-30792.

\end{document}